\documentclass[twocolumn]{aastex62}
\usepackage{tabularx}
\usepackage{graphicx}
\usepackage{amssymb,amsmath,bm,tensor,braket}
\usepackage{xcolor}
\usepackage[varg]{txfonts}
\usepackage{enumerate}
\usepackage{mathtools}
\usepackage[utf8x]{inputenc}
\usepackage[normalem]{ulem}
\usepackage{hyperref}
\usepackage{graphicx}
\usepackage[capitalize]{cleveref}
\crefname{equation}{Eq.}{Eqs.}
\Crefname{equation}{Equation}{Equations}
\usepackage[normalem]{ulem}

\newcommand{\rmd}{{\rm d}}

\graphicspath{{./}{figures/}}
\received{January 1, 2020}
\revised{January 1, 2020}
\accepted{January 1, 2020}
\submitjournal{ApJ}
\shorttitle{Conservation Laws in Gravity}
\shortauthors{Miao, Zhao et al.}
\begin{document}

\title{\Large Tests of conservation laws in post-Newtonian gravity with
binary pulsars}

\correspondingauthor{Lijing Shao}
\email{lshao@pku.edu.cn}
\correspondingauthor{Norbert Wex}
\email{wex@mpifr-bonn.mpg.de}

\author[0000-0003-1185-8937]{Xueli Miao}
\affil{School of Physics and State Key Laboratory of Nuclear Physics and
Technology, Peking University, Beijing 100871, China}

\author[0000-0002-9233-3683]{Junjie Zhao}
\altaffiliation{The first two authors contributed equally to the work.}
\affiliation{School of Physics and State Key Laboratory of Nuclear Physics and
Technology, Peking University, Beijing 100871, China}

\author[0000-0002-1334-8853]{Lijing Shao}
\affiliation{Kavli Institute for Astronomy and Astrophysics, Peking
University, Beijing 100871, China}
\affiliation{Max-Planck-Institut f\"ur Radioastronomie, Auf dem H\"ugel 69,
D-53121 Bonn, Germany}

\author[0000-0003-4058-2837]{Norbert Wex}
\affiliation{Max-Planck-Institut f\"ur Radioastronomie, Auf dem H\"ugel 69,
D-53121 Bonn, Germany}

\author[0000-0002-4175-2271]{Michael Kramer}
\affiliation{Max-Planck-Institut f\"ur Radioastronomie, Auf dem H\"ugel 69,
D-53121 Bonn, Germany}
\affiliation{Jodrell Bank Centre for Astrophysics, The University of
Manchester, Oxford Road, Manchester M13 9PL, United Kingdom}

\author[0000-0002-5660-1070]{Bo-Qiang Ma}
\affiliation{School of Physics and State Key Laboratory of Nuclear Physics
and Technology, Peking University, Beijing 100871, China}
\affiliation{Collaborative Innovation Center of Quantum Matter, Beijing,
China}
\affiliation{Center for High Energy Physics, Peking University, Beijing
100871, China}

\begin{abstract}
General relativity is a fully conservative theory, but there exist other
possible metric theories of gravity. We consider non-conservative ones with
a parameterized post-Newtonian (PPN) parameter, $\zeta_2$. A non-zero
$\zeta_2$ induces a self-acceleration for the center of mass of an
eccentric binary pulsar system, which contributes to the second time
derivative of the pulsar spin frequency, $\ddot{\nu}$. In our work, using
the method in \citet{Will:1992ads}, we provide an improved analysis with
four well-timed, carefully-chosen binary pulsars. In addition, we extend
Will's method and derive $\zeta_2$'s effect on the third time derivative of
the spin frequency, $\dddot{\nu}$. For PSR~B1913+16, the constraint from
$\dddot{\nu}$ is even tighter than that from $\ddot{\nu}$. We combine
multiple pulsars with Bayesian inference, and obtain an upper limit,
$\left|\zeta_{2}\right|<1.3\times10^{-5}$ at 95\% confidence level,
assuming a flat prior in $\log_{10} \left| \zeta_{2}\right|$. It improves
the existing bound by a factor of three. Moreover, we propose an analytical
timing formalism for $\zeta_2$. Our simulated times of arrival with
simplified assumptions show binary pulsars' capability in limiting
$\zeta_{2}$, and useful clues are extracted for real data analysis in
future. In particular, we discover that for PSRs B1913+16 and
J0737$-$3039A, $\dddot{\nu}$ can yield more constraining limits than
$\ddot{\nu}$.
\end{abstract}

\section{Introduction} 
\label{sec:intro}

In particle physics, conservation laws play an important role. They are
default methods to analyze the scattering problem of particles, and have helped
scientists in discovering new particles, e.g. neutrons and
neutrinos~\citep{Chadwick:1932ma,Cowan:1992xc}. In the field of
gravitation, conservation laws do not apply to all metric theories of
gravity. Already at the first post-Newtonian level, they are violated in
some gravity theories. For example, extending the Brans-Dicke theory,
\citet{Smalley:1975ry} constructed a class of gravitational theories with
consistent field equations but non-zero divergence of the energy-momentum
tensor \citep[see also][]{Rastall:1973nw}.

At the first post-Newtonian order, the degree of violation of conservation
laws is expressed via some specific parameterized post-Newtonian (PPN)
parameters~\citep{Will:2018bme}. In order to test the post-Newtonian
gravity, we can bound the PPN parameters in a generic way. These bounds can
be translated to theory parameters afterwards \citep{Will:2018bme}.

For a fully conservative theory, the energy, linear momentum, and angular
momentum are conserved, and PPN parameters satisfy $\alpha_{1} = \alpha_{2}
= \alpha_{3} = \zeta_{1} = \zeta_{2} = \zeta_{3} = \zeta_{4} = 0$
\citep{Will:2018bme}. For semi-conservative theories, the energy and linear
momentum are conserved, but a preferred frame is allowed to exist, which
breaks the symmetry of local Lorentz invariance (LLI) for the gravitational
interaction. In these theories, PPN parameters satisfy $ \alpha_{3} =
\zeta_{1} = \zeta_{2} = \zeta_{3} = \zeta_{4} = 0$ \citep{Will:2018bme}.
Empirically, the best constraints on the other two PPN parameters, namely
$\alpha_1$ and $\alpha_2$, respectively come from observations of the
small-eccentricity binary pulsar PSR~J1738+0333~\citep{Damour:1992ah,
Freire:2012mg, Shao:2012eg} and two solitary millisecond pulsars,
PSRs~B1937+21 and J1744$-$1134~\citep{Nordtvedt:1987ads, Shao:2013wga}.
Non-conservative theories violate the energy-momentum conservation laws,
and one or more of $\{\alpha_{3},\zeta_{1},\zeta_{2},\zeta_{3},\zeta_{4}\}$
will be non-zero~\citep{Will:2018bme}.

In this work, we consider a class of theories with a non-zero $\zeta_2$.
\citet{Smalley:1975ry} explicitly showed that $\zeta_2$ could indeed appear
non-zero in gravitational theories with non-vanishing divergence of the
energy-momentum tensor. \citet{Will:1976zz, Will:1992ads} discovered that
in these theories, the center of mass of an eccentric binary system
possesses an energy-momentum-violating self-acceleration. Therefore, we can
use binary pulsars to test the PPN parameter $\zeta_2$.

In 1974, \cite{Hulse:1974eb} discovered the first binary pulsar system,
PSR~B1913+16. This system provided a verification of the existence of
gravitational-wave radiation for the first time~\citep{Taylor:1979zz}. The
orbital period decay rate, $\dot{P}_{b}$, of this system is consistent with
the predicted value from general relativity (GR). Such an observational
fact can be used to test the foundation of gravity. For example, $\dot P_b$
was used to test if the graviton is massless~\citep{Finn:2001qi,
Miao:2019nhf}. The high-precision results for PSR~B1913+16 benefit from an
extremely accurate measurement technique, the so-called {\it pulsar
timing}. Pulsar timing models the times of arrival (TOAs) of pulses emitted
from a pulsar and determines timing parameters to a high precision via
fitting to a timing formula~\citep{Taylor:1992kea}. With its help, we can
use binary pulsar systems to perform various tests of
gravity~\citep{Stairs:2003eg, Wex:2014nva, Shao:2016ezh}.

Until now, GR has passed all tests with flying colors~\citep{Will:2014kxa}.
However, it is still important to look for gravity theories beyond GR and
also to test GR more and more precisely \citep{Berti:2015itd}. If there
exist non-conservative effects in the gravitational interaction, there
could be a self-acceleration for the center of mass of an eccentric binary
system~\citep{Will:1976zz, Will:1992ads}. It leads to abnormal changes in
the observed pulsar spin and orbital periods. Therefore, we can use binary
pulsar systems to perform gravitational tests and constrain the
corresponding PPN parameters. Note that such constraints on PPN
parameters are in the strong-field regime, because neutron stars are
strongly self-gravitating objects. Tests with binary pulsars are therefore sensitive to strong-field modifications of the weak field PPN parameter $\zeta_2$.\footnote{In the absence of non-perturbative phenomena, one could think of an expansion in terms of the compactnesses ${\cal C}_i$ of the pulsar and its companion: $\zeta_2 = \zeta_2^{\rm PPN} + a_i{\cal C}_i + a_{ij}{\cal C}_i{\cal C}_j + \dots$.} The latest bound on $\zeta_{2}$ was
obtained by \cite{Will:1992ads}. He used the second time derivative of the
spin period (namely $\ddot P$) of PSR~B1913+16, and limited $\zeta_{2}$ to
be smaller than $4\times10^{-5}$ at 95\% confidence level (C.L.). Worth to
note that, the value of $\ddot P$, used by \citet{Will:1992ads}, was
evidently obtained from unpublished work by J.\ H.\ Taylor and colleagues,
but more recent data give a less constraining bound
\citep{Weisberg:2016jye}. This might have been caused by the existence of
red noise, which over a long timing baseline can mimic higher-order spin
period derivatives. It makes the bound over-optimistic by more than an
order of magnitude. Therefore, we shall treat the limit in
\citet{Will:1992ads} as an optimistic one.

In this paper, we perform an improved analysis of binary pulsars to
constrain the strong-field counterpart of the PPN parameter $\zeta_2$. First, using the method
in~\cite{Will:1992ads}, we utilize four carefully chosen binary pulsar
systems to constrain $\zeta_{2}$, including PSR~B1913+16 with the updated
data. The analysis depends on the value of the longitude of periastron,
$\omega$. We attempt to include the effect from the relativistic periastron
advance of the orbit, which renders the value of $\omega$ as a linear
function of time, and the time variation appears to be significant for some
systems. We adopt two methods to constrain $\zeta_{2}$ with different
choices of $\cos\omega(t)$, where $\omega(t)$ is the time-dependent
longitude of periastron. In both cases, the best bound with an individual binary
pulsar is from PSR~B2127+11C~\citep[][Ridolfi et al. in
preparation]{Jacoby:2006dy},
\begin{equation}
   \left|\zeta_{2}\right| \lesssim 3\times10^{-5} \quad (\rm 95\% \, C.L.)\,.
\end{equation}
It is already tighter than the previous best bound
obtained from PSR~B1913+16 \citep{Will:1992ads}.

In addition, we extend Will's method and derive the relation between the
third time derivative of the spin frequency, $\dddot{\nu}$, and
$\zeta_{2}$. Notice that, in this work, we will use time derivatives of the
pulsar spin frequency, $\ddot\nu$ and $\dddot\nu$, instead of time
derivatives of the pulsar spin period, $\ddot P$ and $\dddot P$, that were
used by \citet{Will:1992ads}. These two approaches are equivalent after
properly accounting for the chain rule in taking time derivatives. 
In pulsar timing, the use of frequency derivatives yields a simpler description of the pulsar's spin phase versus time, and it is widely adopted.
In our analysis, the values of $\dddot{\nu}$ are attainable for
PSRs~B1913+16 and B1534+12, and they are used to bound $\zeta_{2}$.
Interestingly, for PSR~B1913+16, the constraint from $\dddot{\nu}$ is even
tighter than that from $\ddot{\nu}$.

With a coherent approach of the Bayesian inference, we combine individual
bounds from four binary pulsars. We obtain a combined bound with a prior
uniform in $\log_{10} \left| \zeta_2 \right|$,
\begin{equation}
    \left|\zeta_{2}\right|< 1.3 \times 10^{-5} \quad (\rm 95\% \, C.L.)\,.
\end{equation}
It improves \citet{Will:1992ads}'s limit by a factor of three.

Moreover, we develop, for the first time, a timing formula that includes $\zeta_2$.
We use it to investigate the capability of
limiting $\zeta_{2}$ from individual binary pulsars. We simulate TOAs for each
pulsar with the effect of $\zeta_{2}$ included, and investigate the ability
to constrain $\zeta_{2}$. 
If the effect of $\zeta_2$ is smaller than the sensitivity of a system to it (which depends on the orbital characteristics and TOA accuracy), then the $\zeta_2$ can not be measured.
It is shown that, if
there were only white Gaussian noise, as it is in our simulation, the
Hulse-Taylor pulsar PSR~B1913+16 would achieve the tightest upper limit,
due to its long observational span and small timing residuals. However, the
existence of red noise in data will deteriorate the test in reality.
Interestingly, in this new method, we find as well that for PSRs B1913+16
and J0737$-$3039A, the third time derivative of the spin frequency,
$\dddot{\nu}$, can yield a stronger constraint on $\zeta_{2}$ than
$\ddot{\nu}$.

The paper is organized as follows. In the next section, we briefly review
the binary dynamics with the PPN parameter $\zeta_{2}$. In
Section~\ref{sec:will}, using an improved method of \citet{Will:1992ads},
we calculate the $\zeta_2$ bounds from four binary pulsars individually.
Then, by including all four pulsars in the Bayesian inference, we obtain a
combined bound on $\zeta_2$. In Section~\ref{sec:toa}, we develop a new
timing formula and simulate TOAs with the contribution of $\zeta_{2}$. We
show the ability to limit $\zeta_{2}$ from different pulsars based on their
current observational characteristics. Though the simulations are
oversimplified with white Gaussian noise and uniform cadence, they still
provide some useful clues for future analysis with real data. We summarize
our results in Section~\ref{sec:discussion}.

\section{Binary pulsars with PPN $\zeta_2$} 
\label{sec:dyn}

For non-conservative gravity theories with the PPN parameter $\zeta_2$,
there exists a self-acceleration for the center of mass of a binary
system~\citep{Will:1976zz, Will:1992ads}. The extra acceleration vector
reads,
\begin{equation}
    \boldsymbol{a}_{\rm cm}(t)=\frac{\zeta_{2}}{2} c T_{\odot} m_{c}
    \frac{q(q-1)}{(1+q)^2}
    \left(\frac{2 \pi}{P_{b}}\right)^{2} \frac{e}{\left(1-e^{2}\right)^{3 / 2}} \hat{\mathbf{e}}_{\mathrm{P}}(t)\,,
    \label{eqn:massacc}
\end{equation}
where, $m_{p}$ and $m_{c}$ are respectively the masses for the pulsar and
its companion star in the Solar unit; $q\equiv m_{p}/m_{c}$ is the mass
ratio; $P_{b}$ is the orbital period, and $e$ is the orbital eccentricity;
$\hat{\mathbf{e}}_{\mathrm{P}}(t)$ is a unit vector directed from the
center of mass of the system to the point of periastron of the pulsar; $c$
is the speed of light, and $T_{\odot}\equiv GM_{\odot}/c^{3} \simeq
4.9254909\, {\rm \mu s}$ with $M_{\odot}$ denoting the Solar
mass~\citep{Mamajek:2015jla}. In addition to $\zeta_2$, other PPN
parameters may as well contribute to Eq.~(\ref{eqn:massacc}), but they have
been constrained tightly~\citep{Will:2014kxa, Will:2018bme}. In this work,
we focus on the impact of $\zeta_2$.

The self-acceleration (\ref{eqn:massacc}) for the center of mass of a
binary system signals a violation of post-Newtonian energy-momentum
conservation~\citep{Will:1976zz, Will:1992ads}. It leads to a {\it
changing} Doppler shift between the Solar system and the binary pulsar
system, due to a uniform rotation of $\hat{\mathbf{e}}_{\mathrm{P}}(t)$
caused by the relativistic periastron advance. This effect changes the {\it
observed} pulsar spin and orbital frequencies~\citep{Will:1976zz,
Will:1992ads}. In this work, we only consider the $\zeta_{2}$-induced
change in the spin frequency. This change can be described via $\dot{\nu} /
\nu \approx -\boldsymbol{a}_{\rm cm}\cdot \hat{\bf n}/c$, where $\hat{\bf
n}$ is a unit vector along the line of sight to the binary. The change of
the pulsar spin frequency, $\dot \nu$, is generally degenerate with its
intrinsic spindown value \citep{Lorimer:2005misc}. Therefore, we turn to
the change of the second time derivative, $\ddot \nu$, which is caused by
the changing orientation of $\hat{\mathbf{e}}_{\mathrm{P}}(t)$ (thus,
$\boldsymbol{a}_{\rm cm}$), due to the advance of the periastron in GR for
a relativistic binary.

If we assume that $\ddot \nu$ is entirely from the contribution of
$\boldsymbol{\dot{a}}_{\rm cm}$, the relation between $\ddot \nu$ and
$\zeta_{2}$ is,
\begin{equation}
  \frac{1}{\nu}\frac{\rmd^{2} \nu}{\rmd t^{2}}= -\frac{\mathcal{A}_{2}}{c}
  \cos\omega\frac{\rmd \omega}{\rmd t}\,,
\label{eq:ddotf}
\end{equation}
where
\begin{equation}
 \mathcal{A}_{2} \equiv-\frac{\zeta_{2}}{2} c T_{\odot} \left(\frac{2
 \pi}{P_{b}}\right)^{2}\frac{q(q-1)}{(1+q)^{2}} \frac{
 e}{\left(1-e^{2}\right)^{3 / 2}}m_{c} \sin i\,.
 \label{eq:A2}
\end{equation}
For convenience, we make use of the mass function,
\begin{equation}
 m_{c}\sin{i}= \left({\cal G} M_{\odot} \right)^{-1/3} \left(\frac{2\pi
 m}{P_{b}}\right)^{2/3}a_{p}\sin{i}\,.
\end{equation}
In above three equations, $\omega$ is the longitude of periastron, $m
\equiv m_p + m_c$ is the total mass of the binary system in the unit of
$M_\odot$, $i$ is the orbital inclination and $(a_{p}/c)\sin{i} \equiv
x_{p}$ is the projected semi-major axis of the pulsar orbit. Though the
{\it effective} gravitational constant ${\cal G}$ in principle could
deviate from its Newtonian counterpart $G$, in particular in the presence of strongly self-gravitating bodies, they were constrained to be close from several pulsar
systems \citep{Shao:2016ezh}. Therefore, we safely take ${\cal G} = G$ in
our calculation as an approximation. By inverting Eq.~(\ref{eq:ddotf}), it
is straightforward to see that, if $e$ and $\dot\omega$ are large enough,
and $|q - 1|$ does not vanish, a limit of $\zeta_{2}$ can be obtained using
the measurement of $\ddot \nu$.

The second time derivative of spin frequency (\ref{eq:ddotf}) is equivalent
to Eq.~(3) in \citet{Will:1992ads} for the second time derivative of spin
period, after dropping negligible higher-order terms. \citet{Will:1992ads}
chose the second time derivative of the spin period of PSR~B1913+16 to
constrain $\zeta_{2}$, and he got a tight bound, $|\zeta_{2}|< 4 \times
10^{-5}$ at 95\% C.L.. In this work, we largely follow the spirit of
\citet{Will:1992ads}, while making several improvements to his method.

Besides the $\ddot \nu$ test, we extend Will's work to bound $\zeta_{2}$
with the third time derivative of the pulsar spin frequency, $\dddot{\nu}$,
and investigate what kind of constraint can be obtained from it. We derive
the relation between $\dddot{\nu}$ and $\zeta_{2}$ by using the same method
in the Appendix of \citet{Will:1992ads}. After dropping higher-order
contributions, we get,
\begin{equation}
  \frac{1}{\nu}\frac{{\rm d}^{3}\nu}{{\rm d}t^{3}}=\frac{\mathcal{A}_{2}}{c}
  \sin\omega \left(\frac{\rmd \omega}{\rmd t}\right)^{2}\,.
	\label{eq:dddotf}
\end{equation}

\section{New limits on $\zeta_2$} 
\label{sec:will}

In this section, we apply Will's method~\citep{Will:1992ads} to the latest
published parameters of four binary pulsars, in order to place updated
bounds on the PPN parameter $\zeta_{2}$. In~\cref{subsec:selection}, we
show our strategy to choose binary pulsar systems with high figure of
merit. In~\cref{ssubsec:individ_bound} the latest parameters of four binary
pulsars are made use of to constrain $\zeta_{2}$ from individual binary pulsars.
We stack them to obtain a combined bound on $\zeta_{2}$ with the Bayesian
inference in~\cref{ssubsec:combined_bound}.

\begin{deluxetable*}{lllll}
\tablecaption{Relevant parameters for PSRs~B2127+11C \citep{Jacoby:2006dy},
B1534+12 \citep{Fonseca:2014qla}, B1913+16 \citep{Weisberg:2016jye}, and
J1756$-$2251 \citep{Ferdman:2014rna}. Their $\ddot{\nu}$ was obtained from
pulsar timing data directly. The $\ddot \nu$ for PSR~J1756$-$2251 was
provided by R.~Ferdman (private communication) using data in
\citet{Ferdman:2014rna}, and we use updated $\ddot \nu$ and masses for
PSR~B2127$+$11C from Ridolfi et al. (in preparation). Binary masses were
obtained assuming the validity of GR. The contribution to $\ddot{\nu}$ from
the magnetic dipole braking, $\ddot{\nu}^{\rm dipole}$, is calculated
assuming a braking index $n=3$. Parenthesized numbers represent the
1-$\sigma$ uncertainty in the last digit(s) quoted. \label{tab:PSRs1}}
\tablehead{&
\colhead{PSR~B2127$+$11C}
& \colhead{PSR~B1534$+$12}
& \colhead{PSR~B1913$+$16}
& \colhead{PSR~J1756$-$2251}
}
 \startdata
  Reference time, $t_0$ (MJD) & $50000$ & $52077$ & $52984$ & $53563$\\
  Observational span, $T^{\rm obs}$ (yr) & $\sim12$ & $\sim 22$ & $\sim31$
  & $\sim9.6$\\
  Spin frequency, $\nu$ (Hz) & $32.755422697308(11)$~~~&
  $26.38213277689397(11)$~~~ & $16.940537785677(3)$~~~ &
  $35.1350727145469(6)$\\
  First derivative of $\nu$, $\dot{\nu}$ $({\rm s^{-2}})$ & $-5.35160(3)
  \times 10^{-15}$ & $-1.686097(2) \times 10^{-15}$ &
  $-2.4733(1)\times10^{-15}$ & $-1.256079(3) \times 10^{-15}$\\
   Second derivative of $\nu$, $\ddot{\nu}$ $(\rm s^{-3})$ & $2.7(26) \times
   10^{-28}$ & $1.70(11) \times 10^{-29}$ & $1.59(15)\times10^{-26}$ &
   $-2.4(1)\times10^{-27}$ \\
   Third derivative of $\nu$, $\dddot{\nu}$ $(\rm s^{-4})$ & -- & $-1.6(2)
   \times 10^{-36}$ & $-5.0(7)\times10^{-35}$ &
   -- \\
   $\ddot{\nu}^{\rm dipole}$ $(\rm s^{-3})$ & $2.6\times10^{-30}$ & $3.2\times10^{-31}$ & $1.0\times10^{-30}$ & $1.3\times10^{-31}$\\
  Orbital period, $P_{b}$ (day) & $0.33528204828(5)$ & $0.420737298879(2)$&
  $0.322997448918(3)$& $0.31963390143(3)$\\
  Eccentricity, $e$ & $0.681395(2)$ & $0.27367752(7)$ & $0.6171340(4)$&
  $0.1805694(2)$\\
  Projected semi-major axis, $x_{p}$ (lt-s) & $2.51845(6)$ & $3.7294636(6)$
  & $2.341776(2)$& $2.756457(9)$\\
  Longitude of periastron, $\omega$ (deg) & $345.3069(5)$ &
  $283.306012(12)$ & $292.54450(8)$ & $327.8245(3)$\\
  Periastron advance, $\dot{\omega}$ $({\rm deg\,yr^{-1}})$ & $4.4644(1)$ &
  $1.7557950(19)$& $4.226585(4)$& $2.58240(4)$\\
  Pulsar mass, $m_{p}$ $({\rm M_{\odot}})$ & $1.3518(9)$ & $1.3330(2)$ &
  $1.438(1)$ & $1.341(7)$ \\
  Companion mass, $m_{c}$ $({\rm M_{\odot}})$ & $1.3610(9)$ & $1.3455(2)$ &
  $1.390(1)$ & $1.230(7)$\\
  Mass ratio, $q \equiv m_p/m_c$ & $0.993(1)$ & $1.0094(2)$ & $1.0345(10)$
  & $1.090(8)$\\
  Number of TOAs, $N_{\rm TOA}$ & $631$ & $9897$ & $9257$ & $8743$\\
  RMS timing residual, $\sigma_{\rm TOA}$ $({\rm \mu s})$ & $26.0$ & $4.57$
  & $17.5$ & $19.3$
  \\
 \enddata
\end{deluxetable*}


\subsection{Selection of binary pulsars \label{subsec:selection}}

Because of its small timing residuals and a long observational time span
for decades, the Hulse-Taylor pulsar PSR B1913+16 bounded $\zeta_{2}$
tightly~\citep{Will:1992ads}. However, now we have many more relativistic
binary pulsars~\citep{Wex:2014nva, Manchester:2015mda}, which have
potential to provide stronger bounds on $\zeta_{2}$. We make use of the
latest published results of binary pulsars, and update the bound of
$\zeta_{2}$ with improved methods.

We select binary pulsars from the ATNF pulsar
catalog\footnote{\url{https://www.atnf.csiro.au/research/pulsar/psrcat}}~\citep{Manchester:2004bp}
in the hope to include all potential binary pulsars with high figure of
merit. First, we choose binary pulsars who have measured values of $e$,
$\ddot \nu$, $x_{p}$, $\omega$, and $\dot\omega$, which are the parameters
appearing in Eq.~(\ref{eq:ddotf}). Particularly, we select relativistic
binary systems with $\dot \omega>0.03^{\circ}\,{\rm yr}^{-1}$. The eligible
systems are PSRs B2127+11C~\citep{Jacoby:2006dy},
B1534+12~\citep{Fonseca:2014qla}, B1802$−$07~\citep{hlk+04},
J1906+0746 \citep{vanLeeuwen:2014sca}, J0024$−$7204U, J0024$−$7204S,
B0021$−$72H, and B0021$−$72E~\citep{FreireLong}. We obtain approximate
constraints on $\zeta_{2}$ from these systems by setting $\cos\omega=1$ for
a rough estimation. We find that, only PSRs~B2127+11C and B1534+12 have the
potential to constrain $\zeta_2$ to an interesting level. We collect
relevant parameters of PSRs~B2127+11C and B1534+12 in~\cref{tab:PSRs1}. For
PSR~B2127+11C we have listed updated values for $\ddot \nu$ and masses from
Ridolfi et al. (in preparation). We also include PSR~B1913+16
in~\cref{tab:PSRs1}. In the latest publication, \citet{Weisberg:2016jye}
did not report the measurement of $\ddot{\nu}$, but we can access relevant
TOAs and associated online data from~\citet{weisberg_2016_54764}. We use
{\sc TEMPO}\footnote{\url{http://tempo.sourceforge.net}} to obtain the
value of $\ddot \nu$ for PSR~B1913+16. Moreover, we include
PSR~J1756$-$2251 in \cref{tab:PSRs1}, whose $\ddot \nu$ was provided by
R.~Ferdman (private communication), by using the data in
\citet{Ferdman:2014rna}. We also tried to use an approximate formula in
\citet{Shao:2014oha, Shao:2014bfa} to estimate the value of $\ddot \nu$
from the uncertainty of $\dot \nu$. The approximate formula works well for
time derivatives of orbital elements \citep{Shao:2018vul}, but is too
optimistic for the spin parameters. This might be due to the
characteristics of timing uncertainties, in particular in the presence of red timing noise. Therefore, we do not include the
estimation in the calculation.

In a short summary, we have collected four binary pulsars in
Table~\ref{tab:PSRs1} to investigate possible constraints on $\zeta_{2}$.

\subsection{Individual bounds on $\zeta_{2}$ \label{ssubsec:individ_bound}}

\begin{figure}[t]
  \includegraphics[width=9cm]{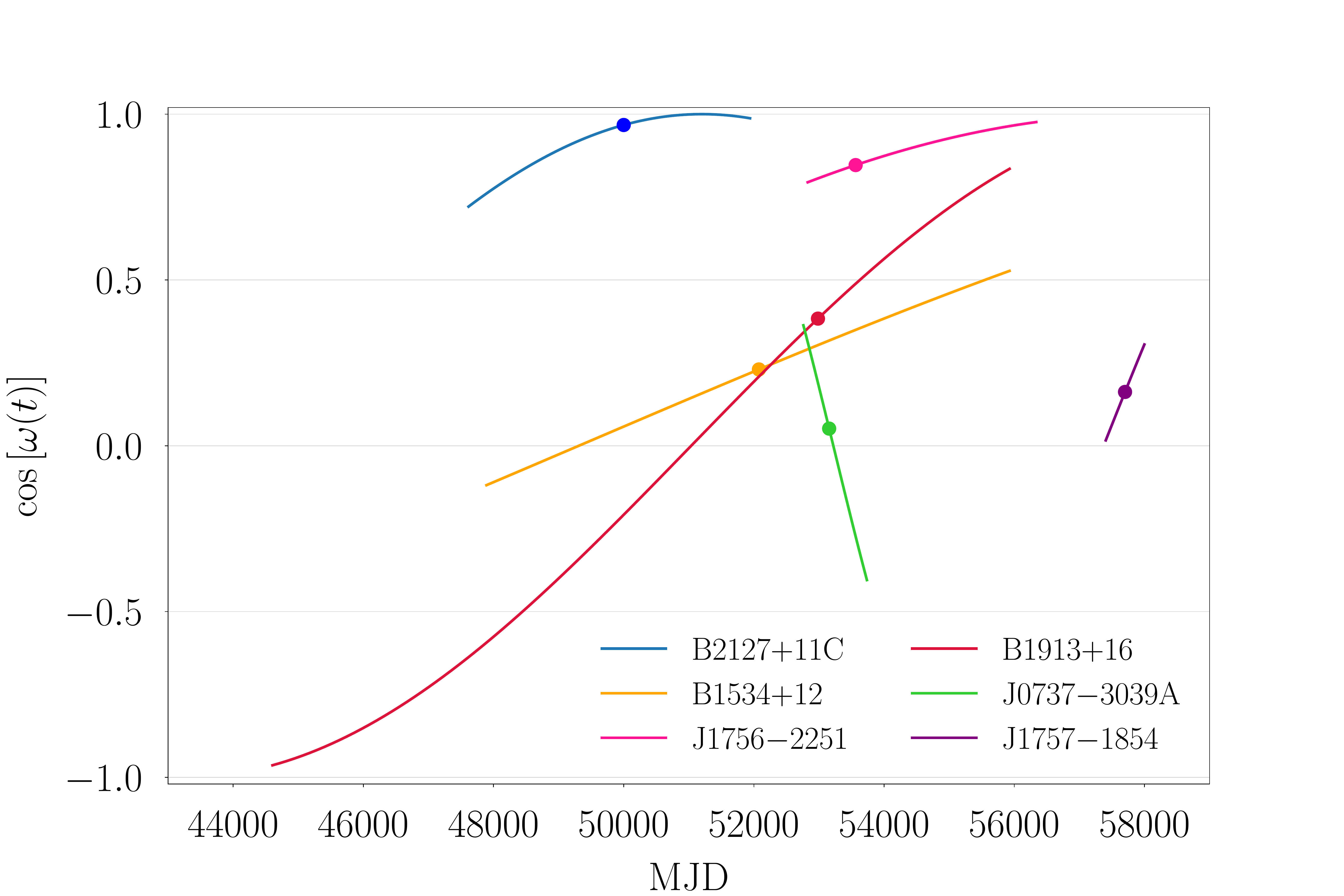}
  \caption{The cosines of the longitude of periastron for six binary
  pulsars in Tables~\ref{tab:PSRs1} and \ref{tab:PSRs2}, during their
  observational spans that were used to derive the timing solution. The
  dots denote the reference epoch for the timing parameters.
  \label{fig:cos}}
  \end{figure}

We use $\ddot{\nu}$ in \cref{eq:ddotf} for four binary pulsars to obtain
individual bounds on $\zeta_{2}$. However, we shall notice that, there
could exist some other effects which contribute to $\ddot{\nu}$. Thus they
should be subtracted before testing $\zeta_2$. It is generally thought that
pulsars have strong dipole magnetic fields. A rotating pulsar leads to the
emission of electromagnetic waves which causes its spindown. The
corresponding $\ddot{\nu}^{\rm dipole}$ can be calculated from the magnetic
dipole braking formula, $\ddot{\nu}^{\rm dipole} = n \dot{\nu}^{2}/\nu$,
where $\nu$ is the spin frequency of the pulsar and $n$ is the so-called
braking index~\citep{Lorimer:2005misc}. For those binary pulsar systems,
the values of $\ddot{\nu}^{\rm dipole}$ are calculated and listed
in~\cref{tab:PSRs1}, assuming $n=3$ for a dominant magnetic dipole braking.
We find that, when compared with the measured $\ddot \nu$, the contribution
from the magnetic dipole braking, $\ddot{\nu}^{\rm dipole}$, is two orders
of magnitude smaller. Therefore, it can be neglected safely when we
constrain the PPN parameter $\zeta_2$ using $\ddot{\nu}$. Other values of
the braking index $n$ give a similar result. Besides the magnetic braking,
the environment of globular clusters and possible nearby masses
\citep{Joshi:1996ci} could also contribute to the time derivatives of the
spin frequency. 
Usually cluster potential will not lead to significant $\ddot\nu$ and $\dddot\nu$. For nearby small-mass objects (e.g. in the PSR~B1620$-$26 system), it is very unlikely to {\it conspire} with $\zeta_2$ to cancel the effect completely.
As for the time derivative of the Galactic acceleration to the
binary pulsar system, we use the data from \citet{Weisberg:2007tb} and
\citet{Weisberg:2016jye} to derive a rough estimation for PSR~B1913+16. We
obtain that such an effect contributes to $\ddot\nu$ at the level of
$10^{-33} \, {\rm s^{-3}}$, far less than the observed $\ddot{\nu} \sim
10^{-26} \, {\rm s^{-3}}$. Therefore, the effect from a time-varying
Galactic acceleration can be ignored as well.

We use the parameters of selected binary systems in Table~\ref{tab:PSRs1}
to place updated bounds on $\zeta_2$ with~\cref{eq:ddotf}. For a parameter
$Y \in\left\{\nu, P_b, e, x_{p}, \dot{\omega}, q, m\right\}$, in the
calculation we take its measured value and associated 1-$\sigma$
uncertainty, $\sigma(Y)$. We generate parameters randomly with a normal
distribution $\mathcal{N} \left[ Y, \sigma^2 (Y) \right]$. In principle,
some of these parameters should change over the observational time span.
For example, the gravitational-wave radiation causes $P_b(t) \simeq
P_b(t_0) + \dot P_b \left(t- t_0\right) $. We have checked that, these
time-varying changes are negligible in putting bounds on $\zeta_{2}$.
Therefore, we directly adopt the values at the reported reference time for
simplicity, with the exception of $\omega$ (see below).

For the key parameter, $\ddot \nu$, to be on the conservative side, we
randomly generate with a normal distribution $\mathcal{N}\left[ 0, \sigma^2
\left( \ddot{\nu}^{\rm upper} \right)\right]$, where, $\sigma
\left(\ddot{\nu}^{\rm upper} \right)$ is the upper limit of $\ddot{\nu}$
from $\zeta_2$. For the four systems in~\cref{tab:PSRs1}, we conservatively
use $\sigma^2 \left(\ddot{\nu}^{\rm upper} \right) ={\ddot{\nu}}^2 +
\sigma^2 \left(\ddot{\nu} \right)$. This treatment is the most
conservative, only assuming that there is no extremely fortuitous
cancellation against $\zeta_2$ with opposite signs from other contributing
sources to $\ddot \nu$.

For some relativistic binary pulsars, the periastron advance is large
enough to be measurable. For example, the Hulse-Taylor pulsar has
$\dot\omega \simeq 4.2^\circ \, {\rm yr^{-1}}$~\citep{Weisberg:2016jye} and
the Double Pulsar has $\dot\omega \simeq 17^\circ \, {\rm
yr^{-1}}$~\citep{Kramer:2006nb}. In order to have a $90^\circ$ change in
$\omega$, the Hulse-Taylor pulsar and the Double pulsar need $\sim 20\,$yr
and $\sim5\,$yr respectively. We note that, in our test there is a
$\cos{[\omega(t)]}$ term in~\cref{eq:ddotf}. It indicates that, differently
from other parameters, the longitude of periastron $\omega$ can vary
greatly over the observational span, thus affecting the $\zeta_2$ test. The
observational spans for binary pulsars in Table~\ref{tab:PSRs1} are of
years to decades, which lead to significant changes in $\omega$ and
$\cos{\left[\omega(t)\right]}$. \Cref{fig:cos} shows the evolution of
$\cos{[\omega(t)]}$ for binary pulsars over their observational span
$T^{\rm obs}$.

For some binary systems, $\cos{[\omega(t)]} $ crosses zero during some
epoch, indicating a loose constraint on $\zeta_{2}$ via
Eq.~(\ref{eq:ddotf}). Consequently, we should treat the value of
$\cos{[\omega(t)]}$ with great caution. In \citet{Will:1992ads},
$\left|\cos{[\omega(t)]}\right|=0.5$ was chosen for PSR~B1913+16. In our
analysis, two different methods for calculating $\zeta_2$ are used:
\begin{itemize}
  \item {\sc Method A}: for each pulsar, we uniformly take the value of
  $\omega(t)$ during its real observational span. The corresponding
  distribution of $\cos{[\omega(t)]}$ is used in our Monte Carlo
  calculation.
  \item {\sc Method B}: for each pulsar, we use the value of
  $\cos{[\omega(t)]}$ at the reference time $t_{0}$, denoted as dots in
  Fig.~\ref{fig:cos}. This reference time is usually chosen to be close to
  the mid-point of the whole observation.
\end{itemize}

Plugging the distribution of $\cos{[\omega(t)]}$ and the distributions of
other relevant parameters with their due uncertainties
into~\cref{eq:ddotf}, we collect the distribution of $\zeta_2$ for
statistical inference. We find that, the distribution of $\zeta_2$ has a
mean value that is very close to zero. We take the symmetric range
enclosing 95\% posteriors as the upper limit of $\zeta_2$ from the
distribution. The corresponding upper limits from different binaries with
{\sc Method A} and {\sc Method B} are given in~\cref{tab:results}.

In our selection of binary pulsars, only two have reported $\dddot{\nu}$ in
their published timing solution. They are PSRs
B1534+12~\citep{Fonseca:2014qla} and B1913+16~\citep{weisberg_2016_54764,
Weisberg:2016jye}. In particular, we have used the {\sc TEMPO} software to
get the value of $\dddot\nu$ for PSR~B1913+16 from their published online
data \citep{weisberg_2016_54764}. The values of $\dddot{\nu}$ for these two
pulsars are listed in~\cref{tab:PSRs1}. Similar to the previous
calculation, we randomly generate a normal distribution for $\dddot \nu$
with $\mathcal{N}\left[ 0, \sigma^2 \left( \dddot{\nu}^{\rm upper} \right)
\right]$, where $\sigma^2 \left( \dddot{\nu}^{\rm upper} \right) =
\dddot{\nu}^2 + \sigma^2 \left(\dddot{\nu} \right)$. We utilize {\sc
Method~A} and {\sc Method~B} to obtain the value of $\omega(t)$. The upper
limits of $\zeta_2$ at 95\% C.L. are obtained from the probabilistic
distributions and are given in~\cref{tab:results} as well.

Let us turn the attention to results in Table~\ref{tab:results}. In {\sc
Method~A}, the tightest constraint is from PSR~B2127$+$11C,
$\left|\zeta_{2}\right|<3.1\times10^{-5}$ (95\% C.L.). It is slightly
better than Will's result. For PSR B1913+16, our bound is
$\left|\zeta_{2}\right|<1.2\times10^{-3}$ from {\sc Method A}, which is
about 30 times looser than the previous limit \citep{Will:1992ads}. The
dominant reason for a worse result is the value of $\ddot P \equiv
-\ddot\nu/\nu^2 + 2\dot \nu^2 / \nu^3$. \citet{Will:1992ads} used an
unpublished value, $\ddot{P} = 4\times10^{-30} \, {\rm s^{-1}}$, which is
actually more than an order of magnitude smaller than the recently
published value in \cite{weisberg_2016_54764}, $\ddot{P} =
5.6\times10^{-29} \, {\rm s^{-1}}$. Such a difference is not surprising in
the presence of red-noise processes, since the value of $\ddot \nu$
determined from timing data is likely to be affected by the time-span of
the available data set \citep{hlk+04}. There is no simple relationship
between the degree of variation of the actually measured $\ddot \nu$ value
and the length of the timing data set, as higher order spin-frequency
derivatives may be required additionally in order to describe the measured
arrival times adequately \citep{hlk10}. Red-noise processes also known as
``timing noise'' or ``spin-noise'', affecting the measured spin-frequency
derivatives, are common for young pulsars and may be related to the
recovery from rotational instabilities known as ``glitches'' \citep{hlk10}
or changes in the pulsar magnetosphere \citep{lhk+10}. Red spin-noise is
also expected to be common in recycled pulsars \citep{sc10}, but evidently
at a much smaller level \citep{hlk10,lhk+10}. Not many studies have
especially addressed the case of so-called ``mildly recycled'' pulsars
which are studied here. However, the magnitude of $\ddot \nu$ values
presented in Table~\ref{tab:PSRs1} matches the expectation and general
trends across the pulsar population \citep{hlk10}.

Another reason is that we uniformly take the
value of $\omega(t)$ during the corresponding observational span. In
\cref{fig:cos}, we notice that the value of $\cos{[\omega(t)]}$ of PSR
B1913+16 goes through zero, which could lead to a significant portion of
samples of $\zeta_2$ with nearly no constraint. Therefore, the distribution
of $\zeta_{2}$ has a very long tail. We have checked that, the limits at
68\% C.L. are much smaller than half of the limits at 95\% C.L., thus
showing evidence of the non-Gaussian long tails in the posterior
distribution. We have encountered a similar situation of long-tailed
distributions for the test of PPN parameter $\alpha_2$ \citep{Shao:2012eg}.
The result from {\sc Method B} also shows the evidence that we could get a
better constraint, $\zeta_{2}<8.4\times10^{-4}$, when we use the value of
$\cos{[\omega(t)]}$ at the reference time $t_{0}$ for PSR B1913+16, when
$\cos{[\omega(t)]}$ is different from zero. For these reasons, we treat the
original limit from \citet{Will:1992ads} as an optimistic one, and ours
more conservative.

\begin{deluxetable}{p{2cm}p{0.9cm}p{1.8cm}p{1.8cm}}
  \tablecaption{The bounds on the absolute value of $\zeta_2$ from
  individual binary pulsar systems at 95\% C.L.. We list the results in the
  order of $\zeta_{2}$ bounds in {\sc Method A}. The second column gives
  the quantity in deriving the constraint. The combined bounds from a
  Bayesian analysis can be found in
  Eqs.~(\ref{eq:1st:linearprior}--\ref{eq:2nd:logprior}).
  \label{tab:results}}
  \tablehead{Pulsar & ~ & {\sc Method A} & {\sc Method B}}
  \startdata
    B2127+11C & $\ddot \nu$ & $3.1\times 10^{-5}$ & $2.9 \times 10^{-5}$\\
    J1756$-$2251 & $\ddot \nu$ & $1.7\times 10^{-4}$ & $1.8 \times10^{-4}$ \\
    B1534+12 & $\ddot \nu$ & $4.5 \times 10^{-4}$ & $8.1 \times 10^{-5}$\\
    B1913+16 & $\dddot \nu$ & $1.2 \times 10^{-3}$ & $8.4 \times 10^{-4}$\\
    B1534+12 & $\dddot \nu$ & $1.9 \times 10^{-3}$& $1.9 \times 10^{-3}$ \\
    B1913+16 & $\ddot \nu$ & $4.1 \times 10^{-3}$ & $1.5 \times 10^{-3}$\\
 \enddata
\end{deluxetable}

In {\sc Method~A}, we have made an improvement in treating the changing
$\omega(t)$. But for PSRs B1913+16 and B1534+12, their
$\cos{[\omega(t)]}\approx0$ during some epoch. So {\sc Method~B} can
provide a stronger limit than {\sc Method~A} for the two pulsars. For PSRs
B2127+11C and J1756$-$2251, their values of $\cos{[\omega(t)]}$ stayed away
from zero during their observational spans, so {\sc Method~A} and {\sc
Method~B} give similar results. In {\sc Method~B}, the best constraint
comes from PSR B2127+11C, $\zeta_{2}<2.9\times10^{-5}\,(\rm 95\% \, C.L.)$.
It is very close to the corresponding limit from {\sc Method~A}.

The bounds from $\dddot{\nu}$ are also listed in Table~\ref{tab:results}
for PSRs B1913+16 and B1534+12. It is worth noting that, PSR B1913+16 can
provide a better bound from $\dddot{\nu}$ than from $\ddot{\nu}$. It
indicates that, at least for some pulsars, 
$\dddot{\nu}$ can offer a stronger bound on $\zeta_2$. Therefore, if
observers could publish $\ddot{\nu}$ and $\dddot\nu$ parameters in the
future, it will help to test the non-conservativeness of gravity theories.

\subsection{A combined bound on $\zeta_{2}$ \label{ssubsec:combined_bound}}

We can stack the posteriors from four pulsars to obtain a combined limit on
$\zeta_2$ via Monte Carlo simulations within the Bayesian framework, as
suggested in the context of \citet{DelPozzo:2016ugt}. In the Bayesian
inference, given a prior, the posterior distribution of $\zeta_2$ can be
inferred with data, ${\cal D}$, and a hypothesis, ${\cal H}$. We use the
Bayes' theorem,
\begin{equation}
	P\left(\zeta_2 | {\cal D}, {\cal H}, {\cal I} \right) = \int
	\frac{P\left( {\cal D} | \zeta_2, \bm{\Xi}, {\cal H}, {\cal I}\right)
	P\left( \zeta_2, \bm{\Xi} | {\cal H},{\cal I}\right)} {P\left({\cal D}
	| {\cal H}, {\cal I} \right)} {\rmd} \bm{\Xi} \,, \label{eqn:bayes}
\end{equation}
where ${\cal I}$ denotes all other relevant knowledge, and $\bm{\Xi}$
collectively denotes all other unknown parameters. In the equation,
$P\left(\zeta_2 | {\cal D}, {\cal H}, {\cal I} \right)$ is an updated
(marginalized) posterior distribution of $\zeta_2$, $ P\left( {\cal D} |
\zeta_2, \bm{\Xi}, {\cal H}, {\cal I}\right) \equiv \mathcal{L} $ is the
likelihood function, $P\left( \zeta_2, \bm{\Xi} | {\cal H},{\cal I}\right)$
is the prior on parameters $\in \left\{\zeta_2, \bm{\Xi}\right\}$, and
$P\left({\cal D} | {\cal H}, {\cal I} \right)$ is the model evidence.

Before investigating the bound on $\zeta_2$, we construct the logarithmic
likelihood function,
\begin{equation}
  \ln \mathcal{L} = - \frac{1}{2} \sum \left[ \frac{\ddot \nu}{ \sigma
  \left(\ddot{\nu}^{\rm upper} \right) } \right]^2 -\frac{1}{2} \sum \left[
  \frac{\dddot \nu}{ \sigma \left(\dddot{\nu}^{\rm upper} \right) }
  \right]^2 \,, \label{eq:likelihood}
\end{equation}
where the $\ddot \nu$ and $\dddot \nu$ in the numerator are the
contributions from $\zeta_2$ [cf. Eq.~(\ref{eq:ddotf}) and
Eq.~(\ref{eq:dddotf})], and the summations are over eligible systems (see
below). The values of $\sigma \left( \ddot{\nu}^{\rm upper}\right)$ and
$\sigma \left(\dddot{\nu}^{\rm upper}\right)$ were discussed
in~\cref{ssubsec:individ_bound}.

Similarly, we investigate two scenarios. In the first scenario, we use
binary pulsars with measured $\ddot \nu$ and/or $\dddot{\nu}$, and we
utilize {\sc Method~A} for individual binary pulsar systems to deal with
the time-varying $\omega(t)$. In the second scenario, we instead use {\sc
Method~B} to obtain $\omega(t)$.

\begin{figure}[t]
  \includegraphics[width=8.5cm]{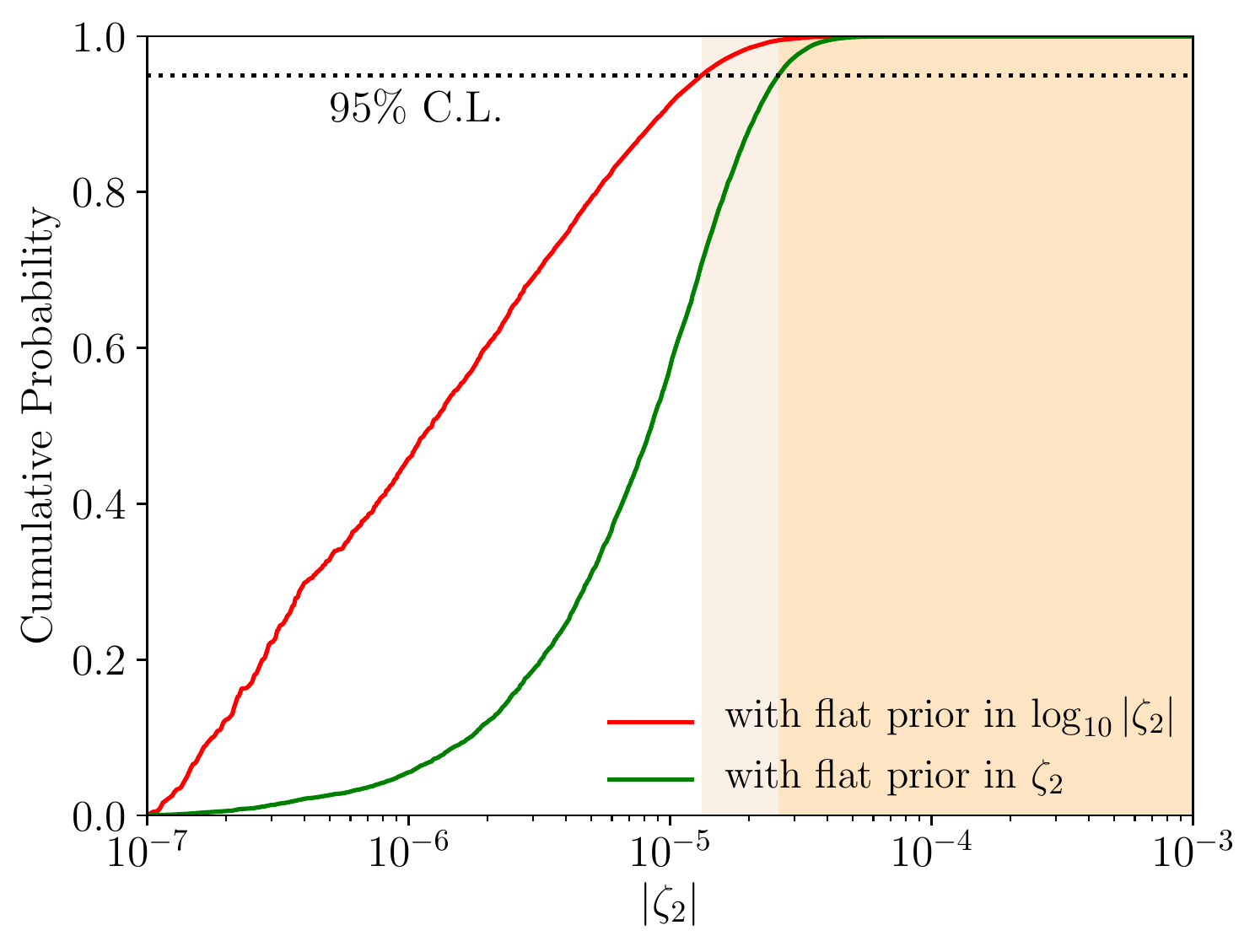}
  \caption{Cumulative posterior distributions with two different choices of
  priors, using the {\sc Method~A} for $\omega(t)$. \label{fig:bayes_wo} }
  \end{figure}

\begin{figure}[t]
  \includegraphics[width=8.5cm]{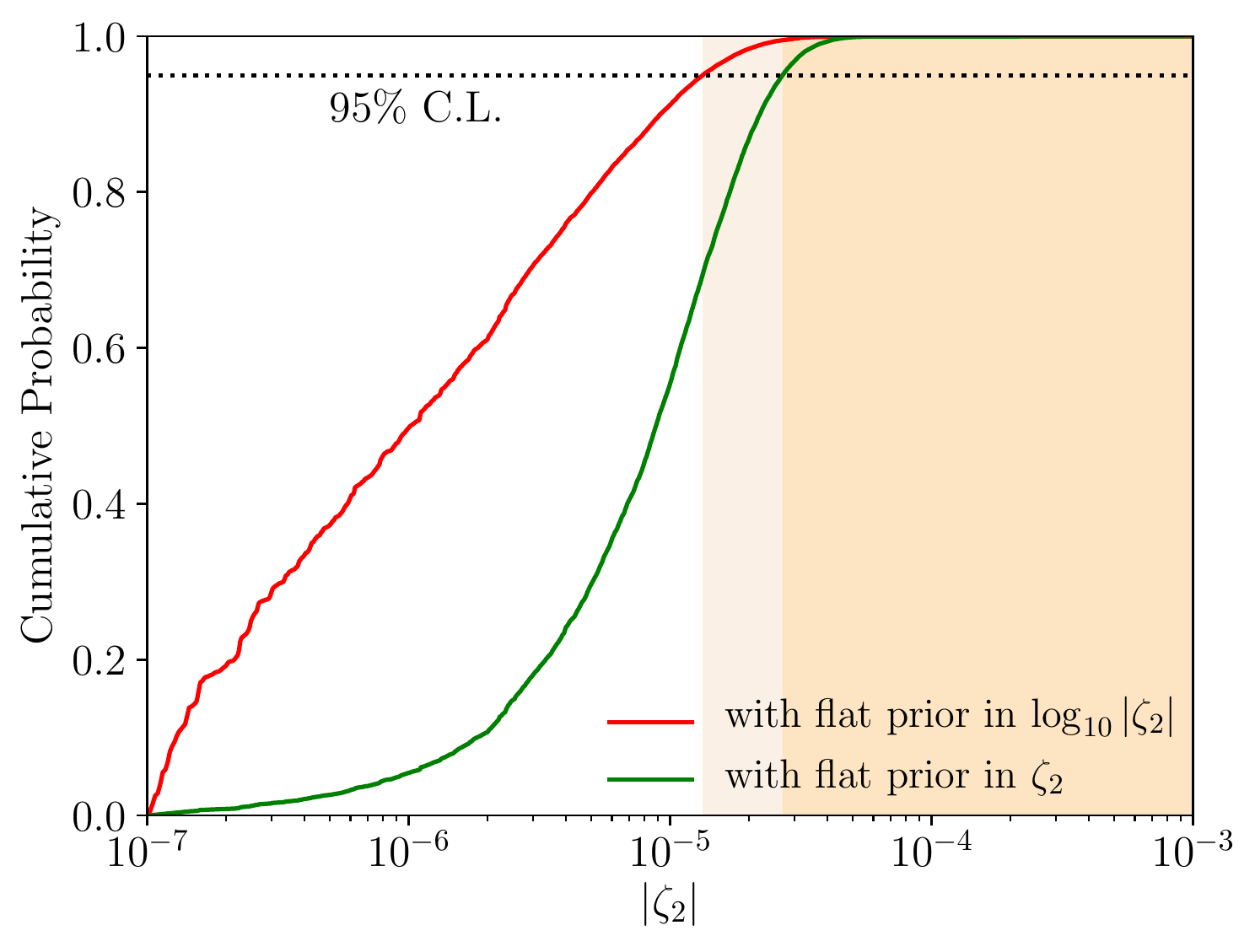}
  \caption{Same as Fig.~\ref{fig:bayes_wo}, using {\sc Method~B} for
  $\omega(t)$. \label{fig:bayes_w}}
\end{figure}

For the two scenarios above, for each we introduce two types of prior
distribution for $\zeta_2$, namely a flat prior on $\log_{10} \left|
\zeta_2 \right|$ in the range $\log_{10} \left| \zeta_2 \right| \in
[-7,\,-3]$, and a flat prior on $\zeta_2$ in the range
$\left|\zeta_2\right| \in [10^{-7},\,10^{-3}]$.

The posterior distributions with different priors are illustrated
in~\cref{fig:bayes_wo} and \cref{fig:bayes_w} for {\sc Method~A} and {\sc
Method~B} respectively. In {\sc Method~A}, the constraints at 95\% C.L.
are,
\begin{align}
  \left| \zeta_{2} \right| &<2.6\times10^{-5} \,, \quad \mbox{with flat
  prior in $\zeta_2$}\,, \label{eq:1st:linearprior}  \\
  \left| \zeta_{2} \right| &<1.3\times10^{-5} \,, \quad \mbox{with flat
  prior in $\log_{10}\left|\zeta_2\right|$}\,. \label{eq:1st:logprior}
\end{align}
The bound in Eq.~(\ref{eq:1st:logprior}) improves the limit in
\citet{Will:1992ads} by three time. In {\sc Method~B}, at 95\% C.L. we have
\begin{align}
  \left| \zeta_{2} \right| &<2.7\times10^{-5} \,, \quad \mbox{with flat
  prior in $\zeta_2$}\,, \\
  \left| \zeta_{2} \right| &<1.3\times10^{-5} \,, \quad \mbox{with flat
  prior in $\log_{10}\left|\zeta_2\right|$}\,. \label{eq:2nd:logprior}
\end{align}
That the two methods provide very close results proves the consistency and
robustness of our approaches. In \cref{tab:results}, except for
PSR~J1756$-$2251, {\sc Method~B} leads to a better constraint than {\sc
Method~A} for individual bounds. But {\sc Method~A} gives similar result
with {\sc Method~B} for the combined bound on $\zeta_{2}$ with two
different types of prior. The reason is related to the long tails of
individual limits in {\sc Method~A}, when $\cos\omega(t)$ crosses zero, as
discussed above. When combining multiple distributions, the long tails are
suppressed.

\section{A full timing model with simulated data}
\label{sec:toa}

\begin{deluxetable}{llll}[t]
\tablecaption{Relevant parameters for
PSRs~J0737$-$3039A~\citep{Kramer:2006nb} and
J1757$-$1854~\citep{Cameron:2017ody}. Their $\ddot \nu$ was not reported in
literature. Masses were obtained assuming the validity of GR. Parenthesized
numbers represent the 1-$\sigma$ uncertainty in the last digit(s)
quoted.\label{tab:PSRs2}}
\tablehead{ &\colhead{PSR~J0737$-$3039A} &\colhead{PSR~J1757$-$1854}
}
 \startdata
  $t_0$ (MJD) & $53156$ & $57701$\\
  $T^{\rm obs}$ (yr) & $\sim2.67$ & $\sim1.6$ \\
  $\nu$ (Hz) & $44.054069392744(2)$ & $46.517617017655(15)$ \\
  $\dot{\nu}$ $({\rm s^{-2}})$ & $-3.4156(1)\times10^{-15}$ &
  $-5.6917(15)\times10^{-15}$ \\
  $\ddot{\nu}^{\rm dipole}$ $(\rm s^{-3})$ & $7.9\times10^{-31}$ &
  $2.0\times10^{-30}$ \\
  $P_{b}$ (day) & $0.10225156248(5)$ & $0.18353783587(5)$ \\
  $e$ & $0.0877775(9)$ & $0.6058142(10)$ \\
  $x_{p}$ (lt-s) & $1.415032(1)$ & $2.237805(5)$ \\
  $\omega$ (deg) & $87.0331(8)$ & $279.3409(4)$ \\
  $\dot{\omega}$ $({\rm deg\,yr^{-1}})$ & $16.89947(68)$ & $10.3651(2)$ \\
  $m_{p}$ $({\rm M_{\odot}})$ & $1.3381(7)$ & $1.3384(9)$ \\
  $m_{c}$ $({\rm M_{\odot}})$ & $1.2489(7)$ & $1.3946(9)$ \\
  $q \equiv m_p/m_c$ & $1.0714(11)$ & $0.9597(9)$\\
  $m$ $(\rm M_{\odot})$ & $2.58708(16)$ & $2.73295(9)$ \\
  $N_{\rm TOA}$ & $131416$ & $3162$\\
  $\sigma_{\rm TOA}$ $({\rm \mu s})$ & $54$ & $36$
 \enddata
\end{deluxetable}

In this section, we investigate the capability to limit $\zeta_{2}$, using
simulations based on the observational characteristics of the chosen binary
pulsars. In Section~\ref{subsec:timing_model}, we derive a new timing model
with a non-zero $\zeta_{2}$. Then in Section~\ref{subsec:simlation} we use
simulation of TOAs to investigate the capability to limit $\zeta_{2}$ by
the pulsar timing techniques. To mimic a usual fitting, we use polynomials
of the time derivatives of the spin frequency at different orders to {\it
absorb} the effect of $\zeta_{2}$.

We use six pulsars as examples. Four of them are given in
Table~\ref{tab:PSRs1}, and additional two are listed in
Table~\ref{tab:PSRs2}. We obtain the sensitivity to $\zeta_2$ with current
observational characteristics of six binary pulsars. For simplicity, for
now we only consider white Gaussian noise in the simulation. Though it can
be over-optimistic compared with the actual situation with red noise
\citep[see e.g.][]{Caballero:2015srj}, our study provides a first
demonstration of the full timing model, and a couple of useful clues for
future investigation (cf. Section~\ref{subsec:analysis}). A simple
validation with real TOAs from PSR~B1913+16 \citep{weisberg_2016_54764}
supports our approach.

\subsection{Timing model with a non-zero $\zeta_2$ \label{subsec:timing_model}}

In pulsar timing, the difference between the predicted TOAs from a best-fit
model and the measured TOAs is called the timing residual. If the timing
residuals do not follow a Gaussian distribution with a mean of zero, it
indicates that there is one or more physical factors which are probably not
taken into account in the fitting~\citep{Lorimer:2005misc}. Therefore, if
$\zeta_{2}$ is large enough, it would lead to systematic deviations in the
timing residuals from a zero-mean Gaussian distribution when it is not
fully degenerate with existing timing parameters.

So far, for the six binary pulsars that we consider, they all nicely fit
with the Damour-Deruelle (DD) timing model which is a phenomenological
model for fully conservative gravity theories and accounts for generic
deviations from GR~\citep{Damour1986, Damour:1991rd}. In other words, there
are no obvious {\it non-conservative} effects of $\zeta_{2}$ in the timing
residuals. It means that, if $\zeta_{2}\neq 0$, the value of $\zeta_{2}$ is
too tiny to be relevant, or its effects are possibly absorbed in other
timing parameters, given the observational uncertainty. Here, we will use
simulated TOAs with effects from $\zeta_{2}$ included directly in the
timing model, to investigate what value of $\zeta_2$ can be visible.

First, we investigate how a non-zero $\zeta_{2}$ will contribute to TOAs.
It will automatically account for the linear-in-time evolution of
$\omega(t)$. Extending the DD timing model~\citep{Damour1985, Damour1986,
Damour:1991rd} we have,
\begin{equation}
  t=T+\Delta_{\rm R}(T)+\Delta_{\rm E}(T)+\Delta_{\rm S}(T)
              +\Delta_{\rm A}(T)+\Delta_{\zeta_{2}}(T)\,,
  \label{eq:ddmodel}
\end{equation}
where $T$ is the proper time of pulsar pulse emission, and $t$ is the
arrival time of pulses at the Solar system barycentre; $\Delta_{\rm R}(T)$
is the Roemer delay, $\Delta_{\rm E}(T)$ is the Einstein delay,
$\Delta_{\rm S}(T)$ is the Shapiro delay, and $\Delta_{\rm A}(T)$ is the
aberration delay \citep[see][for details]{Damour1986}. The last term on the
right-hand side of~\cref{eq:ddmodel}, $\Delta_{\zeta_{2}}(T)$, is the {\it
perturbative} contribution from $\zeta_{2}$.

Now, we try to derive the concrete expression of $\Delta_{\zeta_{2}}(T)$.
As given in Section~\ref{sec:dyn}, due to a violation of the conservation
of energy-momentum via a non-zero $\zeta_{2}$, one has a self-acceleration,
$\boldsymbol{a}_{\rm cm}$, for the center of mass of a binary system
\citep{Will:1992ads}. Its component along the line of sight is
\begin{equation}
  a_r(t)\equiv \hat{\bf n} \cdot \boldsymbol{a}_{\rm cm}(t) =\mathcal{A}_{2}
  \sin[\omega_{0}+\dot{\omega}(t-T_{0})]\,,
\end{equation}
with $T_{0}$ the epoch of periastron, and $\omega_{0}$ the value of the
longitude of periastron at $T_{0}$; ${\cal A}_2$ is given in
Eq.~(\ref{eq:A2}). For simplicity, in the following we will take $T_0=t_0$
which is the reference epoch for the astrometric parameters.

For a pulsar in a binary system, the displacement $z(t)$ along the line of
sight, which is caused by the perturbative effects of $\zeta_{2}$, is
determined via the relation $\ddot{z}(t)=a_r(t)$. After integration, we
obtain
\begin{equation}
z(t)=\frac{\mathcal{A}_{2}}{\dot{\omega}^{2}}\left[\sin \omega_{0}
    +\delta \omega \cos \omega_{0}
    -\sin \left(\omega_{0}+\delta \omega\right)\right]\, , \label{eq:zt}
\end{equation}
where $\delta\omega\equiv\dot{\omega}(t-t_{0})$. We have chosen
$z(t_{0})=\dot{z}(t_{0})=0$ as the initial condition of integration. It is
the most general choice, because other choices could always be absorbed
into parameter redefinition. Due to a non-zero $\zeta_{2}$, the extra time
delay of arrival of pulses can be described by,
\begin{equation} \label{eq:zeta2:timingmodel}
  \Delta_{\zeta_{2}}=z(t)/c\simeq z(T)/c \,.
\end{equation}
Here, the difference between $T$ and $t$ is at higher orders and we will
neglect it. The above equation could be directly applied in pulsar timing
softwares, e.g. {\sc TEMPO}. We have implemented such a timing model with
$\zeta_2$ (see below).

To comply with the tests utilizing $\ddot \nu$ and $\dddot \nu$ that we
mentioned in Section~\ref{sec:will}, we apply Taylor expansion to
$\Delta_{\zeta_{2}}$ with respect to $T-t_{0}$,
\begin{align}
\label{eq:delta}
\Delta_{\zeta_{2}} (T)=&\frac{1}{2} \frac{\mathcal{A}_{2}}{c} \sin \omega_{0}\left(T-t_{0}\right)^{2}+\frac{1}{6} \frac{\mathcal{A}_{2} \dot{\omega}}{c} \cos \omega_{0}\left(T-t_{0}\right)^{3} \nonumber \\ 
&-\frac{1}{24} \frac{\mathcal{A}_{2} \dot{\omega}^{2}}{c} \sin \omega_{0}\left(T-t_{0}\right)^{4}+\ldots\,. 
\end{align}
As we can imagine, $\Delta_{\zeta_{2}}$ would cause the {\it observed} spin
frequency $\nu$ to change as a function of time.

On the other hand, from pulsar astronomy we have the rotational phase of a
pulsar as a Taylor expansion \citep{Lorimer:2005misc},
\begin{align}
    \label{eq:phi} 
    \phi(T)=&\phi_{0}+\nu\left(T-t_{0}\right)+\frac{1}{2} \dot{\nu}\left(T-t_{0}\right)^{2}+\frac{1}{6} \ddot{\nu}\left(T-t_{0} \right)^{3} \nonumber \\
    &+\frac{1}{24} \dddot{\nu}\left(T-t_{0}\right)^{4}+\ldots \,. 
\end{align}

Comparing Eq.~(\ref{eq:delta}) with Eq.~(\ref{eq:phi}), the PPN
$\zeta_2$ will contribute to the time derivatives of the pulsar spin in the
TOA fitting, namely some effects are degenerate. For the extra time delay
that is caused by an {\it apparent} change in the spin frequency, we have
$-\Delta_{\zeta_{2}}=\delta\phi \,P=\delta\phi /\nu$, and
\begin{align}
 \delta\phi=\frac{1}{2} \delta\dot{\nu}\left(T-t_{0}\right)^{2}+\frac{1}{6}
 \delta\ddot{\nu}\left(T-t_{0}\right)^{3}+\frac{1}{24}
 \delta\dddot{\nu}\left(T-t_{0}\right)^{4} +\ldots\,.
\end{align}
Consequently, we have the following relations for the extra time delay
caused by $\zeta_2$,
\begin{align}
  -\frac{\mathcal{A}_{2} \dot{\omega}}{c} \cos \omega_{0}
  &=\delta\ddot{\nu}/\nu\,, \label{eq:toarelation1} \\
 \frac{\mathcal{A}_{2} \dot{\omega}^{2}}{c} \sin \omega_{0}
 &=\delta\dddot{\nu}/\nu\,.
\label{eq:toarelation2}
\end{align}
They are actually equivalent to Eq.~(\ref{eq:ddotf}) and
Eq.~(\ref{eq:dddotf}). If there is only the $\zeta_{2}$ parameter
contributing to $\delta \ddot{\nu}$ and $\delta \dddot{\nu}$,
\cref{eq:toarelation1,eq:toarelation2} can be made use of to test
$\zeta_{2}$. Notice that, we do not consider a possible constraint from
$\delta\dot{\nu}$, because this parameter is much more likely to be
dominated by un-modeled astrophysical processes (for example, by the dipole
radiation of pulsars), thus $\ddot{\nu}$ can provide a more reasonable
constraint than $\dot{\nu}$ \citep{Will:1992ads}. As in
Section~\ref{sec:will}, we will only consider $\ddot \nu$ and $\dddot \nu$
in the following.

\subsection{Simulations and fitting to TOAs \label{subsec:simlation}}

In order to investigate the capability to limit $\zeta_{2}$ with binary
pulsars, we construct simulated TOAs including the effect of $\zeta_2$. For
simplicity, we assume that, the contribution from a non-zero, yet small,
$\zeta_{2}$ does not significantly affect the best-fitting parameters from
pulsar timing. We use the published parameters that were obtained without
considering the time delay effect of $\zeta_2$. These parameters are given
in Tables~\ref{tab:PSRs1} and \ref{tab:PSRs2}. By doing this, we are
assuming that the effects from $\zeta_2$ are perturbatively small. We
consider such an assumption reasonable at the stage of {\it bounding}
$\zeta_2$ instead of {\it measuring} it. In addition, we only assume white
Gaussian noise in our simulation. There could be heterogeneous noise and
significant red noise for some binary pulsars \citep{Caballero:2015srj}.
We feel the assumption of white noise to be optimistic, but still
reasonable, for this {\it demonstrative} study. Further studies can be
conducted to investigate the effects from red noise and more realistic
observational cadence.

For each pulsar, we simulate TOAs, $\{ t_{n} \}_{n=1}^{N_{\rm TOA}}$, where
$N_{\rm TOA}$ is the number of TOAs which were used to derive the actual
pulsar parameters (see the penultimate row in Tables~\ref{tab:PSRs1} and
\ref{tab:PSRs2}). Simplifying the actual observational cadence, here the
$n$-th simulated TOA is expressed as $t_{n} = n \,T^{\rm obs}/N_{\rm TOA}
$, namely, they are chosen to be uniform in the observational span $T^{\rm
obs}$. We utilize their root mean square (RMS) residual, given in the last
row of Tables~\ref{tab:PSRs1} and \ref{tab:PSRs2}, to generate the white
timing noise, $w (t_n)$. We add randomly generated noise to these TOAs. On
top of these fake TOAs, we add the timing delay caused by a non-zero
$\zeta_2$, $\Delta_{\zeta_{2}}(T)$, which is directly obtained from
Eq.~(\ref{eq:zeta2:timingmodel}).

\begin{figure*}[]
  \centering
  \includegraphics[width=14cm]{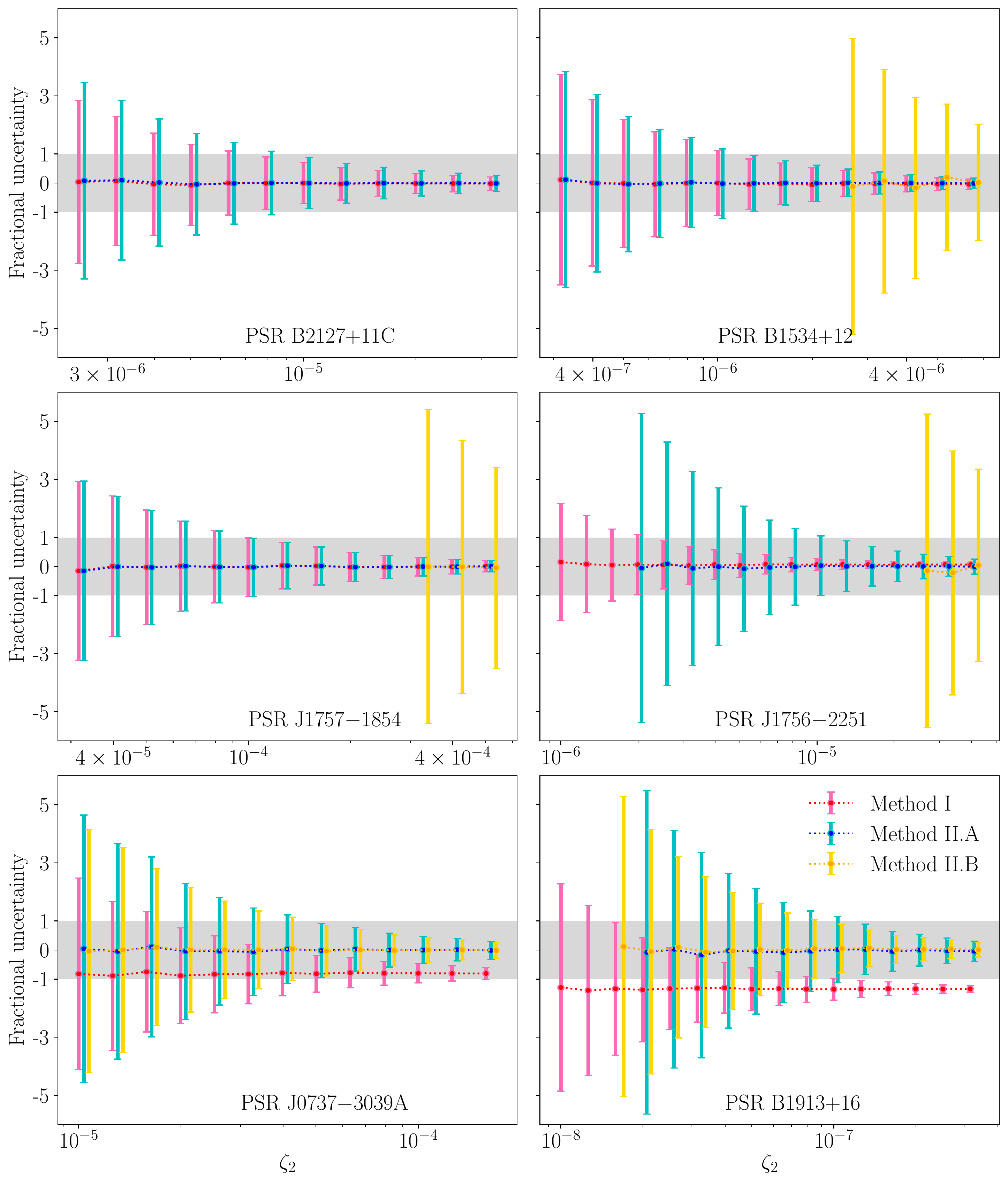}
  \caption{Fractional uncertainties ${\cal R}$ for six binary pulsars from
  three different methods, as a function of $\zeta_2^{\rm input}$ (see text
  for more details). The gray strip is a region which is bound by
  $\sigma({\cal R}) \le 1$. When $\sigma({\cal R})$ is smaller than the
  width of the gray strip, the effect from $\zeta_2$ starts to be
  relevant.} \label{fig:fitting}
  \end{figure*}

Now we try to extract the $\zeta_2$ parameter from simulated timing
residuals. In our investigation, given a realization of $\{ t_{n} \}$ and a
value of $\zeta_{2}^{\rm input}$, we can simulate timing residuals for each
pulsar based on their timing parameters in Tables~\ref{tab:PSRs1} and
\ref{tab:PSRs2}. With these simulated timing residuals, we try to separate
the effect of $\zeta_{2}$, with a simplified timing model. To mimic the
fitting in real situation, we use a polynomial that is expanded with
respect to $T - t_0$,
\begin{equation}
  \label{eqn:fit_func}
  \Delta(T) = \frac{1}{2} \alpha \left(T-t_{0}\right)^{2}+\frac{1}{6} \beta
  \left(T-t_{0}\right)^{3}+\frac{1}{24} \gamma \left(T-t_{0}\right)^{4}
  +\ldots \,,
\end{equation}
where $\alpha$, $\beta$, and $\gamma$ are all fitting parameters. 

When comparing Eqs.~(\ref{eq:delta}) and (\ref{eqn:fit_func}), we observe
the following correspondence: $\alpha = \mathcal{A}_{2} \sin \omega_{0}/c
$, $\beta=\mathcal{A}_{2} \dot{\omega} \cos \omega_{0} /c $ and $\gamma= -
\mathcal{A}_{2} \dot{\omega}^{2} \sin \omega_{0} / c$. These parameters are
treated independently in the fitting, therefore we will put a superscript
to indicate the order of the corresponding polynomial coefficients
hereafter. As we have discussed before, due to the contamination in $\dot
\nu$ \citep{Will:1992ads}, we only utilize the fitting parameters $\beta$
and $\gamma$ to derive bounds on $\zeta_2$. In fact, according
to~\cref{eq:toarelation1,eq:toarelation2}, $\beta$ and $\gamma$ are related
to the usual $\ddot \nu$ and $\dddot \nu$, respectively. As the result of
fitting, the derived $\zeta_{2}$, which we denote as $\zeta_2^{\rm fit}$,
can be obtained from the coefficients at different orders,
$\mathcal{A}^{(3)}_{2}$ and $\mathcal{A}^{(4)}_{2}$.

In order to investigate at which level we will be able to bound
$\zeta_{2}$, we fit the simulated timing residuals with~\cref{eqn:fit_func}
according to the following two schemes. In the first scenario, we fit the
timing residuals with~\cref{eqn:fit_func} up to the {\it third} order,
namely by including $\alpha$ and $\beta$. We derive the value of
$\zeta_{2}^{\rm fit}$ from $\beta$. For convenience, we call it {\sc Method
I}. In the second scenario, we fit the timing residuals
with~\cref{eqn:fit_func} up to the {\it fourth} order, namely by including
$\alpha$, $\beta$, and $\gamma$. Differently from the previous scenario,
now in principle we can obtain two independent bounds on $\zeta_{2}$ from
either $\beta$ or $\gamma$. To make a clear distinction, the methods where
$\zeta_{2}^{\rm fit}$ is derived from $\mathcal{A}^{(3)}_{2}$ and
$\mathcal{A}^{(4)}_{2}$ are named as {\sc Method II.A} and {\sc Method
II.B}, respectively. Worth to note that, in the case that we can contribute
$\ddot \nu$ and $\dddot \nu$ solely to $\zeta_2$, if {\sc Method II.A} and
{\sc Method II.B} give a same value of $\zeta_2$, it represents a way to
{\it detect} $\zeta_2$ other than to {\it bound} $\zeta_2$. But in reality,
it might be difficult to separate other astrophysical contributions to
$\ddot \nu$ and $\dddot \nu$.

Until now, we have introduced how to derive $\zeta_{2}^{\rm fit}$ with a
realization of the white noise $w(t_n)$ and a non-zero $\zeta_2^{\rm
input}$ for a pulsar. Because of the existence of random noise, given a
$\zeta_2^{\rm input}$, $\zeta_2^{\rm fit}$ inherits the randomness.
Therefore we generate a set of realization of $w(t_n)$ and repeat the
simulations and fittings to obtain statistical distributions for
$\zeta_2^{\rm fit}$.

To quantify the difference between $\zeta_{2}^{\rm input}$ and
$\zeta_2^{\rm fit}$, we introduce the fractional uncertainty, ${\cal R}
\equiv \left(\zeta_{2}^{\rm fit}-\zeta_{2}^{\rm
input}\right)/\zeta_{2}^{\rm input}$. For each pulsar, we record the
distribution of $\zeta_{2}^{\rm fit}$, and obtain the distribution of $\cal
R$ from it. The mean of the distribution of $\cal R$ is expected to be zero
for {\it unbiased} fittings. The timing residual from $\zeta_2$ is included
in fake TOAs via Eq.~(\ref{eq:zeta2:timingmodel}), while it is fit via
Eq.~(\ref{eqn:fit_func}). Therefore, intrinsically, we are biased. But as
we will see later, such a bias is not important for most of our binary
pulsars.

The 1-$\sigma$ uncertainty of $\cal R$, denoted as $\sigma({\cal R})$,
describes the pulsar's capability to limit $\zeta_{2}$. When $\sigma({\cal
R}) > 1$, we consider that the effects of $\zeta_2^{\rm input}$ are buried
in noise and cannot be extracted. For the following, we introduce
$\sigma({\cal R}) \leq 1$ as a criterion for detectability. For each
pulsar, we repeat the above processes for multiple values of $\zeta_2^{\rm
input}$, to look for the critical value, which is the smallest value of
$\zeta_2^{\rm input}$ that meets the criterion. For the binary pulsars that
we use, we investigate proper ranges of $\zeta_{2}^{\rm input}$ for each
pulsar individually. With {\sc Method I}, {\sc Method II.A}, and {\sc
Method II.B}, we obtain the value of ${\cal R}$ and $\sigma({\cal R})$
as a function of $\zeta_2^{\rm input}$. The results are illustrated
in~\cref{fig:fitting} for the six chosen pulsars. The critical values of
$\zeta_2^{\rm input}$ for $\sigma({\cal R})=1$ are collected
in~\cref{tab:fitting}.

\begin{table*}
  \caption{Critical values of $\zeta_2$ for six binary pulsars with three
  different methods. Corresponding values of $\ddot{\nu}$ and $\dddot{\nu}$
  are listed next to them. {\sc Method~I} for PSRs~B1913+16 and
  J0737$-$3039A is biased (see Fig.~\ref{fig:fitting}), thus not listed.
  \label{tab:fitting}}
  \centering
  \def\arraystretch{1.5}
  \begin{tabular}{lcccccccc}
  \hline\hline
   & \multicolumn{2}{c}{\textbf{\sc Method I}} &
   \multicolumn{2}{c}{\textbf{\sc Method II.A}} &
   \multicolumn{2}{c}{\textbf{\sc Method II.B}}\\
   & $\zeta_{2}^{\rm crit}$ & $\ddot\nu\,(\rm Hz^{3})$ & $\zeta_{2}^{\rm
   crit}$ & $\ddot\nu\,(\rm Hz^{3})$ & $\zeta_{2}^{\rm crit}$ &
   $\dddot\nu\,(\rm Hz^{4})$\\
  \hline
  B1534+12 & $1.2 \times 10^{-6}$ & $-4.9 \times 10^{-31}$ & $1.2 \times
  10^{-6}$ & $-4.9 \times 10^{-31}$ & $1.2 \times 10^{-5}$ &
  $-2.0\times10^{-38}$\\
  J0737$-$3039A & -- & -- & $4.4 \times 10^{-5}$ & $2.3 \times 10^{-27}$ &
  $3.4 \times 10^{-5}$ & $-3.2\times10^{-34}$ \\
  J1756$−$2251& $2.0 \times 10^{-6}$ & $5.3 \times 10^{-29}$ & $1.1 \times
  10^{-5}$ & $2.9 \times 10^{-28}$ & $1.3 \times 10^{-4}$ &
  $3.1\times10^{-36}$ \\
  J1757$−$1854& $1.0 \times 10^{-4}$ & $-2.8 \times10^{-26}$ & $1.0 \times
  10^{-4}$ & $-2.8 \times 10^{-26}$ & $1.7 \times 10^{-3}$ &
  $-1.7\times10^{-32}$ & \\
  B2127+11C & $7.0 \times 10^{-6}$ & $8.5\times10^{-29}$ & $8.0 \times
  10^{-6}$ & $9.7\times10^{-29}$ & $7.0 \times 10^{-3}$ &
  $5.5\times10^{-35}$ \\
  B1913+16 & -- & -- & $1.0 \times 10^{-7}$ & $2.1\times10^{-30}$ & $8.0
  \times 10^{-8}$ & $9.5\times10^{-39}$\\
  \hline
  \end{tabular}
\end{table*}

\begin{figure}[t]
  \centering
  \includegraphics[width=8cm]{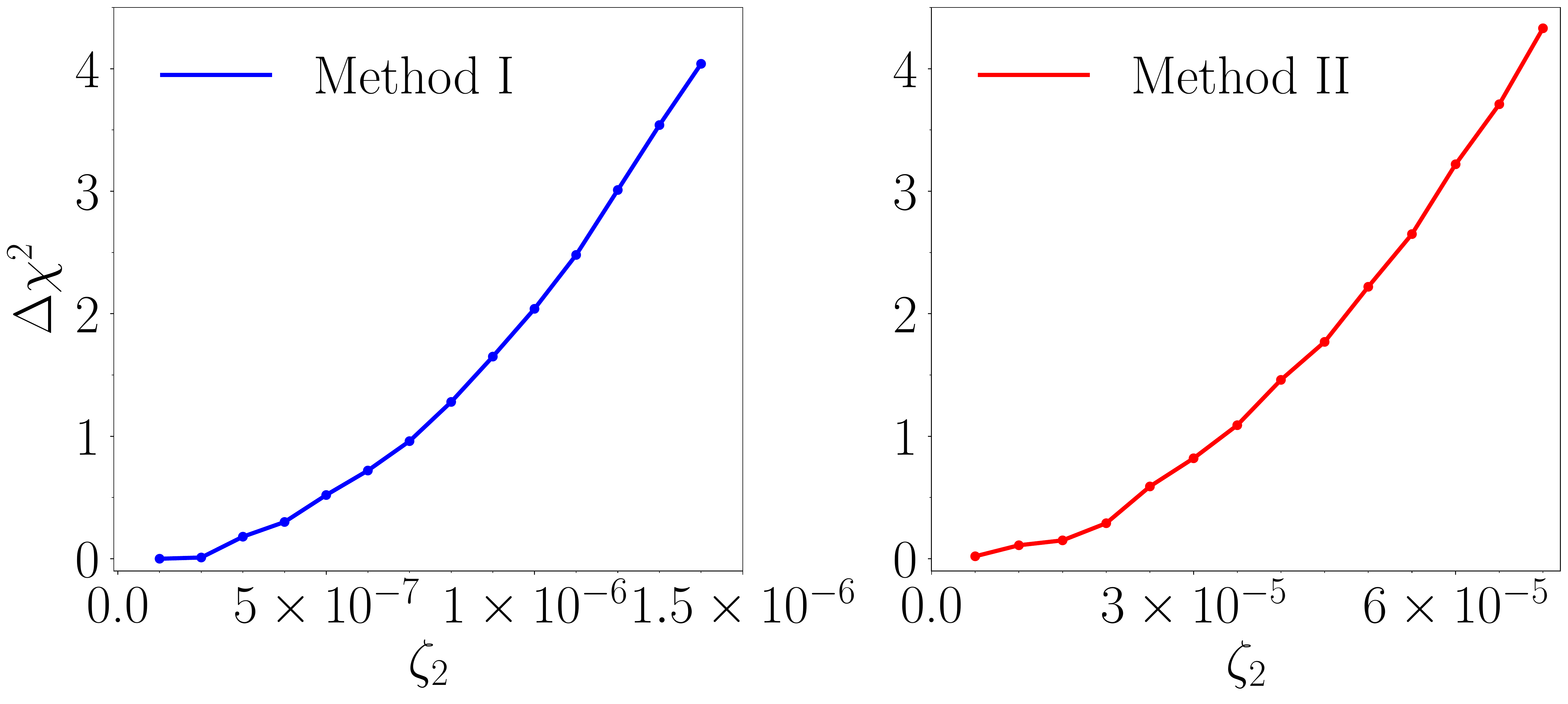}
  \caption{Changes in the $\chi^2$ as a function of $\zeta_2$ for
  PSR~B1913+16.}
  \label{fig:b1913}
  \end{figure}

To have a better sense of implementation, we augment the DD timing model
with the extra time delay in Eq.~(\ref{eq:zeta2:timingmodel}) in the {\sc
TEMPO} software. Using the full timing model in {\sc TEMPO} we have
verified the simplified treatments above. In addition, we apply the new
model to the public data of PSR~B1913+16 \citep{weisberg_2016_54764} from
real observations. Instead of fitting $\zeta_2$ directly, we scan the
values of $\zeta_2$ in appropriate ranges and record the changes in
$\chi^2$. Same as the analysis in this subsection, we freely fit for up to
the second time derivative of the spin frequency in {\sc Method}~I, and up
to the third time derivative of the spin frequency in {\sc Method}~II. Our
results are plotted in Fig.~\ref{fig:b1913}. As we can see, though the
cadence of real data is very different from our simulation, the results are
consistent with our analysis. Worth to note that, in the fitting of
\citet{Weisberg:2016jye}, higher time derivatives of the spin frequency
were used. These parameters are not free in the calculation of
Fig.~\ref{fig:b1913} for simplicity. These higher frequency derivatives
might be caused by red noise. In contrast to this demonstrative work, they
need to be properly accounted for in real data analysis.

\subsection{Discussions \label{subsec:analysis}}

Now, we analyze and discuss the implication of our results in the
\cref{fig:fitting,tab:fitting}. Naturally, as one can see in
Fig.~\ref{fig:fitting}, when $\zeta_2^{\rm input}$ increases, the signal
gets more prominent and the fractional uncertainty in estimating $\zeta_2$
gets smaller. It becomes easier to separate the $\zeta_{2}$ effect from
other noise given by the RMS timing residuals. It indicates a stronger
capability to limit $\zeta_{2}$.

As we also observe in \cref{fig:fitting}, except for PSRs J0737$-$3039A and
B1913+16, when $\zeta_2^{\rm input}$ increases, the mean of $\cal R$
gradually converges to zero with {\sc Method I}. It means that, though the
$\zeta_2$ effects are introduced through the full timing model in
Eq.~(\ref{eq:zeta2:timingmodel}), the fitting using
Eq.~(\ref{eqn:fit_func}) with polynomial coefficients $\alpha$ and $\beta$
are enough to {\it absorb} the residuals, effectively into the
spin-down/spin-up parameters. But for PSRs B1913+16 and J0737$-$3039A, the
$\zeta_2$ effects cannot be absorbed solely with $\alpha$ and $\beta$; the
recovery will be biased if only $\alpha$ and $\beta$ are used.
Nevertheless, if we have included the $\gamma$ coefficient in
Eq.~(\ref{eqn:fit_func}), as shown with {\sc Method II.A} and {\sc Method
II.B}, the $\zeta_2$ effects can be almost totally absorbed.

For a larger value of $\zeta_{2}^{\rm input}$, it is easier to identify the
$\zeta_{2}$ parameter with the pulsar timing data. In our criterion, if
$\zeta_2^{\rm input} < \zeta_2^{\rm crit}$, the effects of $\zeta_2$ are
buried underneath white noise. When $\zeta_{2}^{\rm input}>\zeta_{2}^{\rm
crit}$, we consider that we are able to notice the $\zeta_{2}$ effect via
pulsar timing. Hence, we take $\zeta_2^{\rm crit}$ of a pulsar as its
measure of the capability to limit $\zeta_{2}$. The values of $\zeta_2^{\rm
crit}$ with different methods are listed in Table~\ref{tab:fitting}. It
should be noted that, in the simulation we have used a uniform cadence and
do not consider red noise, so the results from our simulation should be
considered as optimistic estimates. Our main purpose with this section is
to illustrate the timing formalism and simply indicate its possible use in
the future.

Nevertheless, we would like to extract some useful clues for future
studies. According to the different behaviors in the convergence of the
quantity $\cal R$ in Fig.~\ref{fig:fitting}, We divide the six binary
pulsars into three categories for discussions.
\begin{itemize}
  \item In the first category, we have PSRs B1534+12, B2127+11C and
  J1757$-$1854. As shown in~\cref{fig:fitting}, for each of these three
  pulsars, {\sc Method I} and {\sc Method II.A} have a similar capability
  to limit $\zeta_{2}$, while {\sc Method II.B} performs much worse.
  Especially, for PSR~B2127+11C, its $\zeta_{2}^{\rm crit}$ from {\sc
  Method II.B}, $\zeta_{2}^{\rm crit} = 7.0 \times 10^{-3}$, is too large
  and exceeds the plot range of the vertical axis.
  \item In the second category, we have PSR J1756$-$2251. For this pulsar,
  {\sc Method I} provides a tighter result than {\sc Method II.A} and {\sc
  Method II.B}. We notice that in the fitting the coefficients $\beta$ and
  $\gamma$ are highly correlated, which worsens the tests with {\sc Method
  II.A} and {\sc Method II.B}. 
  \item In the third category, we have PSRs B1913+16 and J0737$-$3039A,
  whose central values of ${\cal R}$ from {\sc Method I} deviate
  significantly from zero. Instead, when we use {\sc Method II.A} or {\sc
  Method II.B}, the recovered $\zeta_2$ is not biased from $\zeta_2^{\rm
  input}$ at large. The results urge us to include the contribution from at
  least up to the fourth order of $T-t_0$ in~\cref{eqn:fit_func} when we
  use it to mimic the contribution from $\zeta_2$ [cf.
  Eq.~(\ref{eq:zeta2:timingmodel})] for PSRs B1913+16 and J0737$-$3039A. In
  addition, for these two pulsars, {\sc Method II.B} provides a smaller
  $\zeta_{2}^{\rm crit}$ than {\sc Method II.A}. Therefore, $\dddot \nu$, instead of $\ddot \nu$, will provide a
  stronger test of $\zeta_{2}$ with their observational characteristics.
  This is likely caused by the fact that PSR~J0737$-$3039A has an
  extraordinarily large $\dot{\omega} \simeq 17^\circ \,{\rm yr}^{-1}$,
  while PSR~B1913+16 has been observed from several decades. It also
  confirms our conjecture in~\cref{ssubsec:individ_bound}, that the bound
  on $\zeta_2$ from $\dddot{\nu}$ might be stronger than that from $\ddot{\nu}$
  for some binary pulsar systems. It will be important if we want to apply the
  timing model (\ref{eq:zeta2:timingmodel}) to the new data of Double
  Pulsar (Kramer et al. in preparation).
\end{itemize}

In our simulation for PSR~J0737$-$3039A, we have used the observational
characteristics in \citet{Kramer:2006nb}. The observational span was
$T^{\rm obs}\lesssim 3$\,yr and now the pulsar has been monitored for a
much longer time span. Therefore, it is interesting to investigate its
current ability in bounding $\zeta_2$. We simulate additional TOAs for
$\sim17$\,yr using the same observational cadence and the same level of RMS
noise as in \citet{Kramer:2006nb}. We find that under white Gaussian
noise and uniform observational cadence, it is able to probe $\zeta_2$ at
the level of ${\cal O}\left(10^{-8}\right)$. If Taylor-expanded polynomials
are used, higher-order terms are needed for an unbiased parameter recovery, as its $\omega$ has changed by $\sim200^\circ$ over this time-span of observation.
A publication for a new test of $\zeta_2$ with real decade-long timing data
for PSR~J0737$-$3039A is under plan.

\section{Summary \label{sec:discussion}}

Conservation of energy and momentum is an important property of a gravity
theory. In the PPN framework, the PPN parameter $\zeta_2$ describes a class
of theories that violate the conservation laws \citep{Will:2018bme}. There
are explicit examples for this kind of theories
\citep{Rastall:1973nw,Smalley:1975ry}, where the divergence of the
energy-momentum tensor does not vanish [see e.g.\ Eq.~(9) in
\citet{Smalley:1975ry}]. The PPN parameter $\zeta_2$ is proportional to
this non-vanishing divergence [see Eq.~(43) in \citet{Smalley:1975ry}].
Therefore, a generic bound on $\zeta_2$ can be translated to a bound on the
divergence of the energy-momentum tensor in these theories. A non-zero
$\zeta_2$ leads to characteristic timing behaviors for a pulsar in the
binary, which can be tested via observations \citep{Will:1992ads}. In our
study, we systematically investigate possible bounds on the $\zeta_{2}$
parameter with updated timing solutions for four binary pulsars, utilizing
the time derivatives of their spin frequency.

First, we carefully choose four binary pulsar systems, and for each pulsar
we use the method of \citet{Will:1992ads} to put an individual bound on
$\zeta_2$. To improve the choice of a time-dependent $\omega(t)$, we adopt
two methods. In both methods, PSR B2127+11C provides a stronger bound than
that in \citet{Will:1992ads}. For PSR~B1913+16, the result is about 30
times looser than the previous limit. The loose bound of PSR~B1913+16 is
due to a larger $\ddot{P}$ (or equivalently, $\ddot\nu$) than the one Will
used, as well as the resultant distribution of $\zeta_{2}$ with a
non-Gaussian long tail from the crossing of zero for
$\cos\left[\omega(t)\right]$.

Then, we extend the method in \citet{Will:1992ads} to investigate the
relation between $\dddot{\nu}$ and $\zeta_{2}$. We have access to $\dddot{\nu}$
for PSRs~B1913+16 and B1534+12. From PSR~B1913+16, we obtain a stronger
bound from $\dddot{\nu}$ rather than $\ddot{\nu}$, indicating that $\dddot{\nu}$
could give a tighter bound for some binary pulsars. It is consistent with
simulations in Section~\ref{sec:toa} using a set of completely different
methods. Therefore, we urge observers to publish more frequency derivatives in
order to conduct interesting gravity tests.

To use the maximum potential of an ensemble of pulsars, we derive bounds on
$\zeta_2$ by combining four binary pulsars within the Bayesian framework.
We obtain, using a flat prior for $ \log_{10} \left| \zeta_2 \right| \in
\left[-7,-3\right]$,
\begin{equation}
\left|\zeta_{2}\right|<1.3\times10^{-5}  \quad (\rm 95\% \, C.L.)\,, 
\end{equation}
which improves the result of~\citet{Will:1992ads} by a factor of three.

In addition to using $\ddot \nu$ and $\dddot \nu$, we explore $\zeta_2$'s
direct effect in the timing data. We develop a full timing model that
includes the effects of $\zeta_2$, and implement it in the {\sc TEMPO}
software. We simulate timing residuals for six binary pulsars with their
observational characteristics as input (including RMS timing residuals,
number of TOAs and so on). For each pulsar we obtain their capability to
limit $\zeta_{2}$, represented by a critical value, $\zeta_2^{\rm crit}$.
Using our criterion that the $\zeta_2$ signal is not buried in noise, for
each pulsar we use three methods to derive $\zeta_2^{\rm crit}$, which
represents a lower limit for $\zeta_2$ in order to be detected. When
$\zeta_{2}$ is smaller than $ \zeta_2^{\rm crit}$, it is impossible to
measure $\zeta_{2}$ due to the presence of timing noise. In our
simulation, we have assumed white noise and a uniform observational
cadence. These assumptions have rendered our results quite optimistic ones.
Nevertheless, as the first study, it concludes some useful clues in using
the timing delay from $\zeta_2$ for future real data analysis. For example,
the simulations of PSRs~B1913+16 and J0737$-$3039A show (i) the necessity
to include higher-order time derivatives of the spin frequency if a
polynomial functional is used to mimic the $\zeta_2$ effect, and (ii) the
potential that $\dddot{\nu}$ could provide a tighter bound on $\zeta_2$
other than $\ddot{\nu}$.
As now we have a full timing model, in the future, instead of using frequency derivatives, one can in principle use the full timing model, in combination with red noise modeling to test  the PPN $\zeta_2$ parameter.


\acknowledgments

We thank Clifford Will and Heng Xu for helpful discussions. We are grateful
to Robert Ferdman, Paulo Freire, Vivek Venkatraman Krishnan, and Alessandro
Ridolfi for private communication, and Paulo Freire for carefully reading
the manuscript. This work was supported by the National Natural Science
Foundation of China (11975027, 11991053, 11721303), the Young Elite
Scientists Sponsorship Program by the China Association for Science and
Technology (2018QNRC001), and the Max Planck Partner Group Program
funded by the Max Planck Society. LS, NW and MK acknowledge support from
the European Research Council (ERC) via the ERC Synergy Grant BlackHoleCam
under Contract No.\ 610058. The work was partially supported by the
Strategic Priority Research Program of the Chinese Academy of Sciences
through the Grant No. XDB23010200, and the High-performance Computing
Platform of Peking University.

\software{TEMPO \citep{Nice:2015}}

\bibliography{refs}

\end{document}